\documentclass{PoS}

\def \xmm {$XMM$-$Newton$}

\def \src {IGR\,J16479--4514}

\def \hcm {\hbox {\ifmmode $ atom cm$^{-2}\else atom cm$^{-2}$\fi}}

\def \msun {M$_{\odot}$}

\def \mnras {MNRAS}

\title{Supergiant Fast X--ray Transients: a review}

\ShortTitle{Supergiant Fast X--ray Transients: a review}

\author{\speaker{Lara Sidoli} \thanks{Solicited talk}\\
        INAF, Istituto di Astrofisica Spaziale e Fisica Cosmica, \\
         Via E.\ Bassini 15,   I-20133 Milano,  Italy\\
        E-mail: \email{sidoli@iasf-milano.inaf.it}}

\abstract{Supergiant Fast X-ray Transients are a class of Galactic High Mass X--ray
Binaries with supergiant companions. Their extreme transient X--ray flaring
activity was unveiled thanks to INTEGRAL/IBIS observations. The SFXTs
dynamic range, with X-ray luminosities from 10$^{32}$~erg~s$^{-1}$ to 10$^{37}$~erg~s$^{-1}$,
and long time intervals of low X--ray emission, are puzzling, given that
both their donor star properties and their orbital and spin periodicities seem very
similar to those displayed by massive binaries with persistent X-ray
emission. Clumpy supergiant winds, accretion barriers, orbital geometries and wind anisotropies 
are often invocked to explain their behavior, but still several open
issues remain. A review of the main recent observational results will be 
outlined, together with a summary of the new scenarios proposed to explain
their bright flaring X--ray activity. 
The main result of a long $Suzaku$ observation of the 
SFXT IGR~J16479--4514 with the shortest orbital period is also briefly summarized. 
The observation of the X--ray eclipse in this source allowed us to directly probe
the supergiant wind density at the orbital separation, leading to the conclusion that
it is too large to justify the low X--ray luminosity. A mechanism reducing the accretion
rate onto the compact object is required.
}

\FullConference{An INTEGRAL view of the high-energy sky (the first 10 years) -
9th INTEGRAL Workshop and celebration of the 10th anniversary of the launch\\
                 15-19 October 2012\\
                 Bibliotheque Nationale de France, Paris, France}

\begin{document}

\section{Supergiant Fast X--ray Transients}

Supergiant Fast X--ray Transients (SFXTs) are one of the most spectacular
discoveries obtained by the $INTEGRAL$ satellite (Sguera et al. 2005, 2006; Negueruela et al. 2006).
Their X--ray emission,  characterized by short 
(typically lasting $\sim$100--10,000~s) bright X--ray flares, 
is produced by a compact object (mostly a neutron star, given that  
X--ray pulsations have been detected in about a half of the members of the class),
transiently accreting matter directly from the strong wind of the blue supergiant companion
(e.g. Chaty  2010),
although the mechanism producing the transient X--ray emission is still an open issue.
Indeed, it is difficult to explain why SFXTs, which seem to be massive binary systems 
similar to persistently
accreting High Mass X--ray Binaries (HMXBs),
display so different X--ray properties, showing extreme X--ray transient behavior 
(with X--ray intensity spanning from 2 to 5 order of magnitudes)
instead of persistent X--rays.
Although  their brightest X--ray emission is concentrated into  short flares, the  SFXTs accretion phase
can last a few days (Romano et al. 2007, Sidoli et al. 2009a, Rampy et al. 2009).

$INTEGRAL$ observations, with the large field of view, good angular and spectral resolution, good sensitivity at hard X--rays, 
are particularly suited in discovering new members of the class (ten are now the known SFXTs, with several candidates
with peculiar X--ray flaring emission but no optical identification), 
catching the brightest flares to study the long-term properties 
and source duty cycles (Ducci et al. 2010, Mart\'{\i}nez-N\'u\~{n}ez et al. 2010, Blay et al. 2012), 
and in discovering X--ray periodicities (Drave et al., 2011 and references therein), which are fundamental 
quantities to unveil the nature of these sources.

\begin{figure*}
\centering
\begin{tabular}{cc}
\hspace{-1.5cm}
\includegraphics[height=9.0cm,angle=-270]{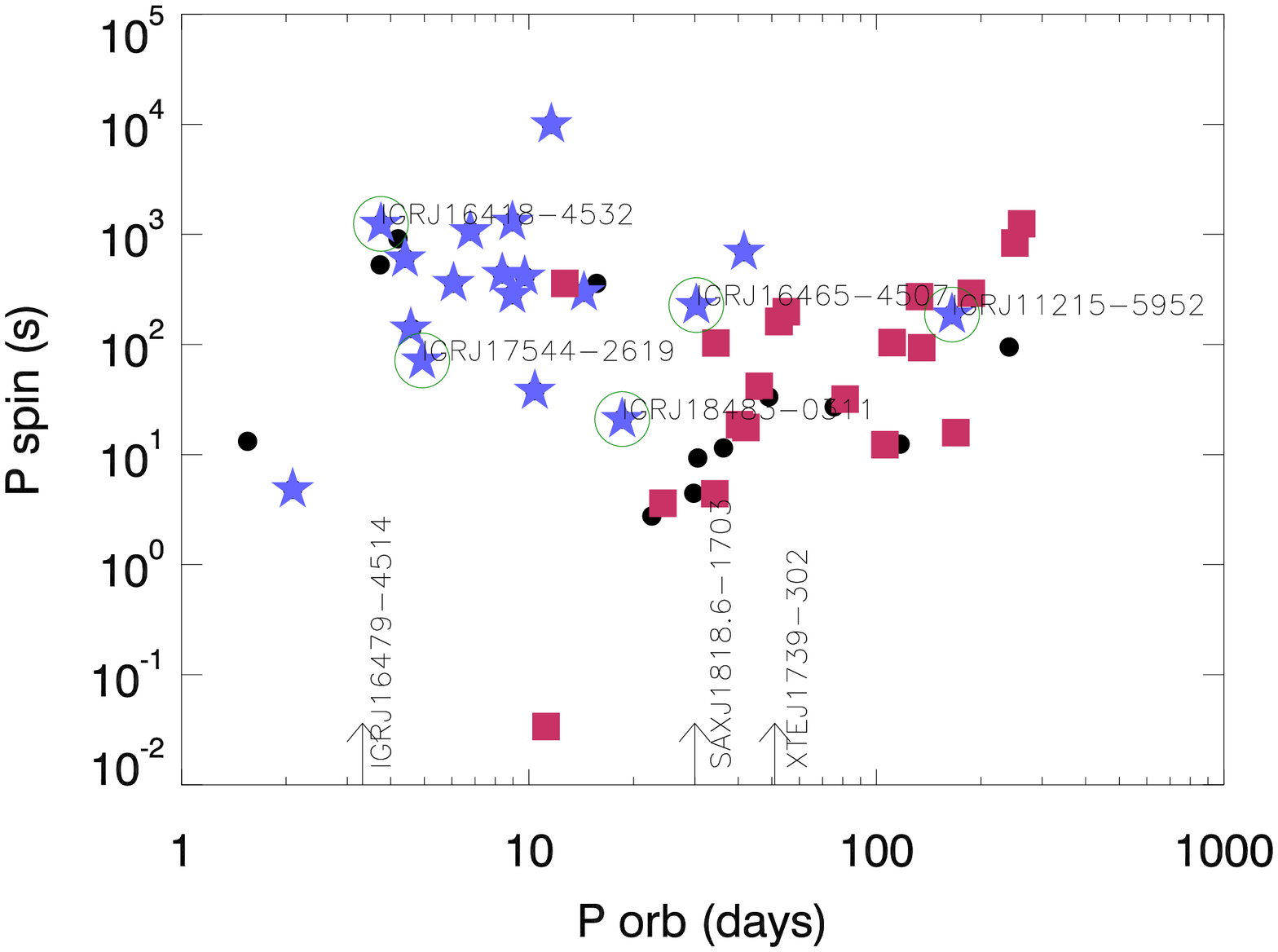} & 
\hspace{-1.0cm}
\includegraphics[height=8.0cm,angle=-270]{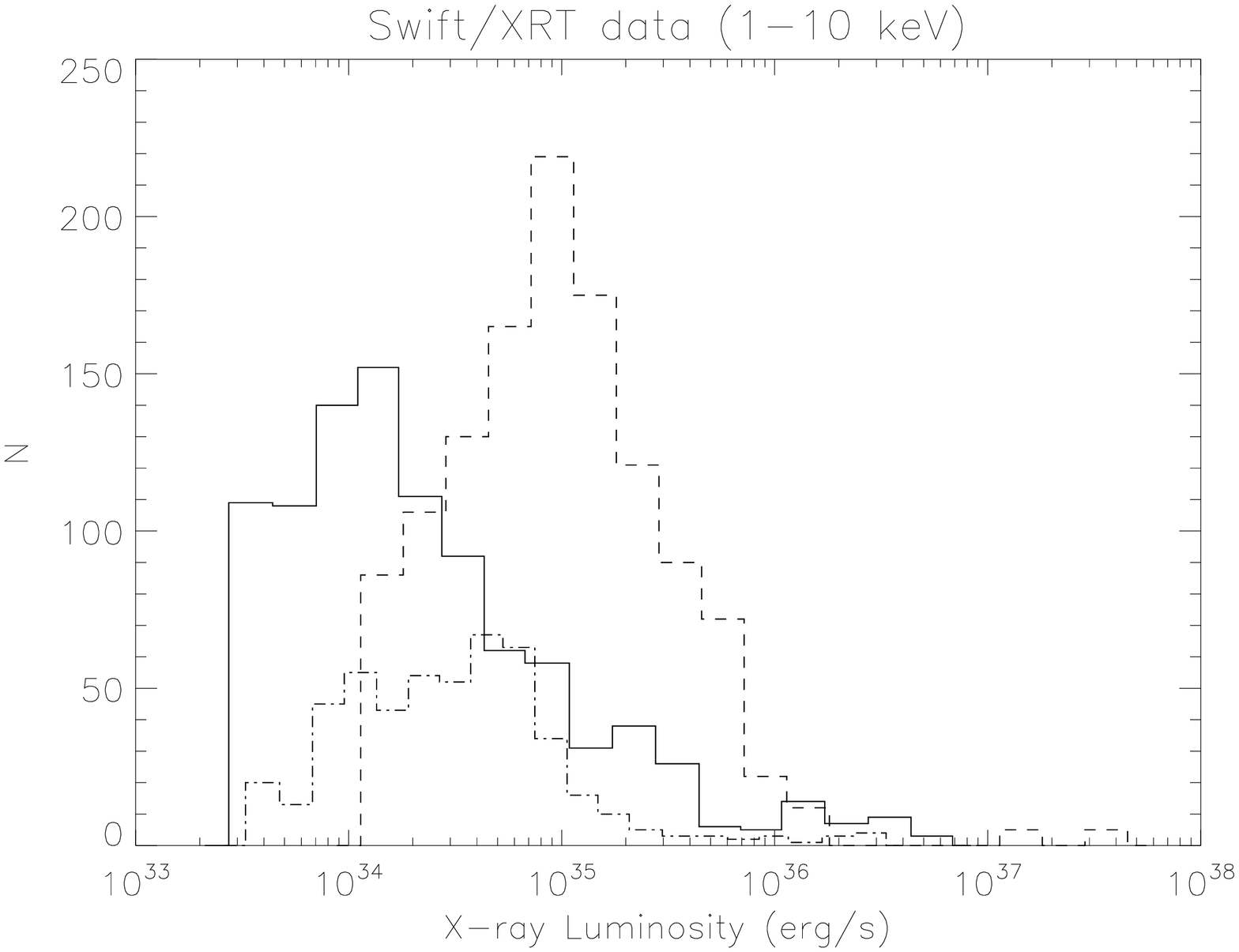} \\
\end{tabular}
\vspace{-0.5cm}
\caption{\scriptsize  
\emph{Left panel}: High Mass X--ray Binary Pulsars located in our Galaxy as they appear in an updated 
version of the Corbet diagram.
{\em  Blue stars} indicate sources  with supergiant companions, while the
{\em  red squares} the pulsars
with Be donors (Liu et al., 2006). 
SFXTs  are indicated by the {\em large circles} around blue stars. 
Arrows mark SFXTs where only the orbital period is known.
\emph{Right panel}: X--ray luminosity distribution calculated for three SFXTs  
(solid line, XTE~J1739-302; dashed line, IGR~J16479-4514; dashed-dotted line, IGR~J17544-2619)
from the 1-10 keV count rates observed by Swift/XRT (count rate distributions taken 
from Romano et al. 2010).
}
\label{lsfig:corbet}
\end{figure*}

The first orbital periodicity in a SFXT was discovered with $INTEGRAL$ from IGR~J11215--5952 (Sidoli et al. 2006), which
displays bright flaring episodes every $\sim$165~days (Sidoli et al. 2007).
Other SFXTs have shown X--ray modulations in $INTEGRAL$ data, associated with their orbital periods 
(Bird et al. 2009, Clark et al. 2009, Drave et al. 2010, Zurita Heras \& Chaty 2009).
Long-term X--ray modulations have been discovered also by other missions, like $Swift$/BAT and $RXTE$/ASM (Corbet et al. 2006, 
Levine et al. 2006, Jain et al. 2009, La Parola et al. 2010), produced by orbital X--ray variability in eccentric and/or eclipsing binaries.
X--ray pulsations have been discovered in a few SFXTs (Fig.~1), 
demonstrating the neutron star nature of the compact object.
These periodicities can be plotted in a Corbet diagram of orbital period versus neutron star spin period (Fig.~\ref{lsfig:corbet},
updated to 2012, October), where SFXTs have been indicated by the large green circles around blue stars.
While persistent HMXB pulsars with supergiant donors (marked by blue stars) mainly concentrate in a region
of the diagram with orbital periods less than 15 days and spin periods around a few hundred seconds, SFXTs 
span across the whole diagram, also lying in regions in-between the wind-fed pulsars with supergiants (in blues) 
and the Be/X-ray transients (in red),
or lying completely inside the region typical for Be/X-ray transients (the SFXT IGR~J11215--5952).

A possibility is that SFXTs represent an evolutionary link between HMXBs with Be companions
and with supergiant donors (Liu et al. 2011; Chaty et al. 2013 these proceedings).
An alternative explanation is that, for SFXTs lying near the Be/XRBs region of this diagram, 
the physical mechanism driving the outbursts is similar
to that producing the Be/XRBs outbursts: the presence of a denser and slower outflowing wind component 
(Sidoli et al. 2007) which triggers the outburst when the  neutron star crosses it.
This could explain the strictly periodic X--ray flares observed from the SFXT IGR~J11215--5952 (Sidoli et al. 2006),
as well as the three peaks  observed in the orbital light curve
of the prototypical SFXT XTE~J1739-302 (Drave et al. 2010).

This is only one of the proposed explanation for the SFXTs behavior. 
Other possibilities are that the compact object accretes matter from an extremely
inhomogeneous  supergiant wind (e.g. in~t Zand 2005; Oskinova et al. 2012), or that
the geometry of the SFXTs orbits are crucial in determining the probability for the 
injestion of massive dense clumps by the neutron star along its orbit.
In persistent systems the neutron star lies always inside the region where the outflowing
clumps are more numerous, while in SFXTs with wider orbits it lies outside it (Negueruela et al. 2008).
$XMM-Newton$ observations of the spectral evolution of the iron line emission
during a bright flare in one SFXT seems to be compatible with the accretion of a single 
dense wind clump (Bozzo et al. 2011).

A result of the $Swift$/XRT two years long campaing on three SFXTs 
(Romano et al. 2010) is the distribution of count rates in
the 1-10 keV energy band. 
From these distributions, I derived the  histogram of the X--ray luminosities (1--10 keV) 
shown in Fig.~\ref{lsfig:corbet} (right panel; Pizzolato \& Sidoli 2013), assuming a power law
spectrum and the distance 
appropriate for each individual SFXT, as reported in Sidoli et al. (2008).
Both the shape of the distributions and their average X--ray luminosity appear to be different in the three SFXTs.
Compared with the luminosity distributions of persistently accreting HMXBs,
where the dynamic range of the X--ray luminosity is limited to about 10 or 100
(only when considering the so called ``off-states'' and the rare ``giant flares'', e.g. Kreykenbohm et al. 2008),
the difference can be due completely to  different properties
of the accreting matter, that is HMXBs
accrete from clumpy winds with a narrow mass distribution, while SFXTs from supergiant 
winds with a much wider clumps' mass distribution (Pizzolato \& Sidoli 2013).

Oskinova et al. (2012) recently studied the clumped stellar winds in supergiant HMXBs.
They performed hydrodynamical simulations and derived the implied X--ray variability, 
mainly due to variability of the wind velocity, the role of which was poorly considered
in previous works. The strong variability in both the density and the velocity
in the structured wind translates (by means of Bondi-Hoyle-Lyttleton wind accretion)
into an extreme X--ray variability up to eight orders of magnitude on short time-scales. 
This X--ray variability is too large and has never been observed in HMXBs (even in SFXTs).
These authors suggested that the details of the accretion process  reduce 
the variability resulting from the stellar wind velocity and density jumps.
Viable possibilities is that clumps are destructed near the accreting neutron star,
or that the density and velocity gradients are smoothed out in
the accretion wake, or that a physical mechanism is at work in {\em damping} the variability.

\begin{figure*}
\centering
\begin{tabular}{cccc}
\hspace{-1truecm}
\includegraphics[height=8.5cm,angle=-270]{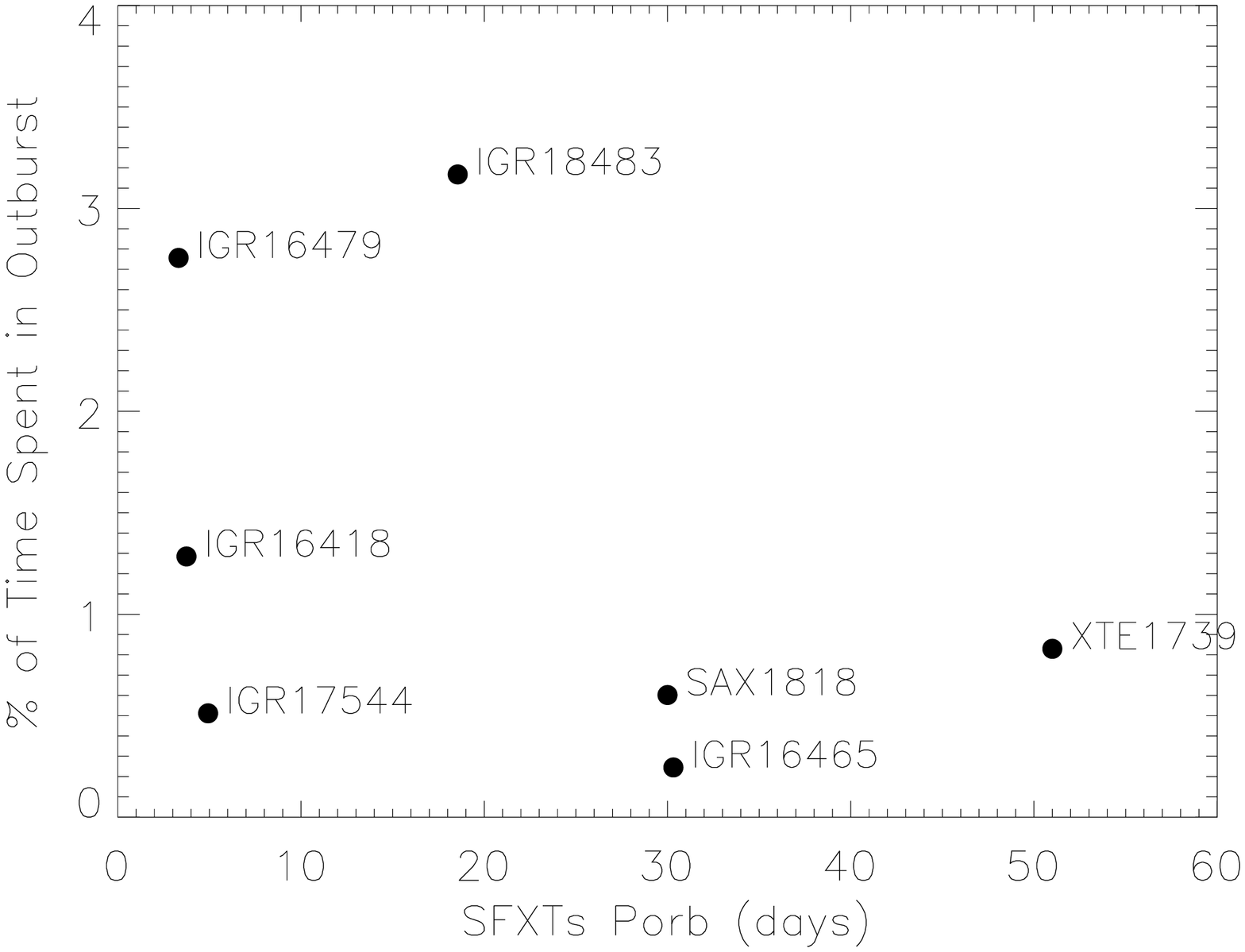} & 
\hspace{-1truecm}
\includegraphics[height=8.5cm,angle=-270]{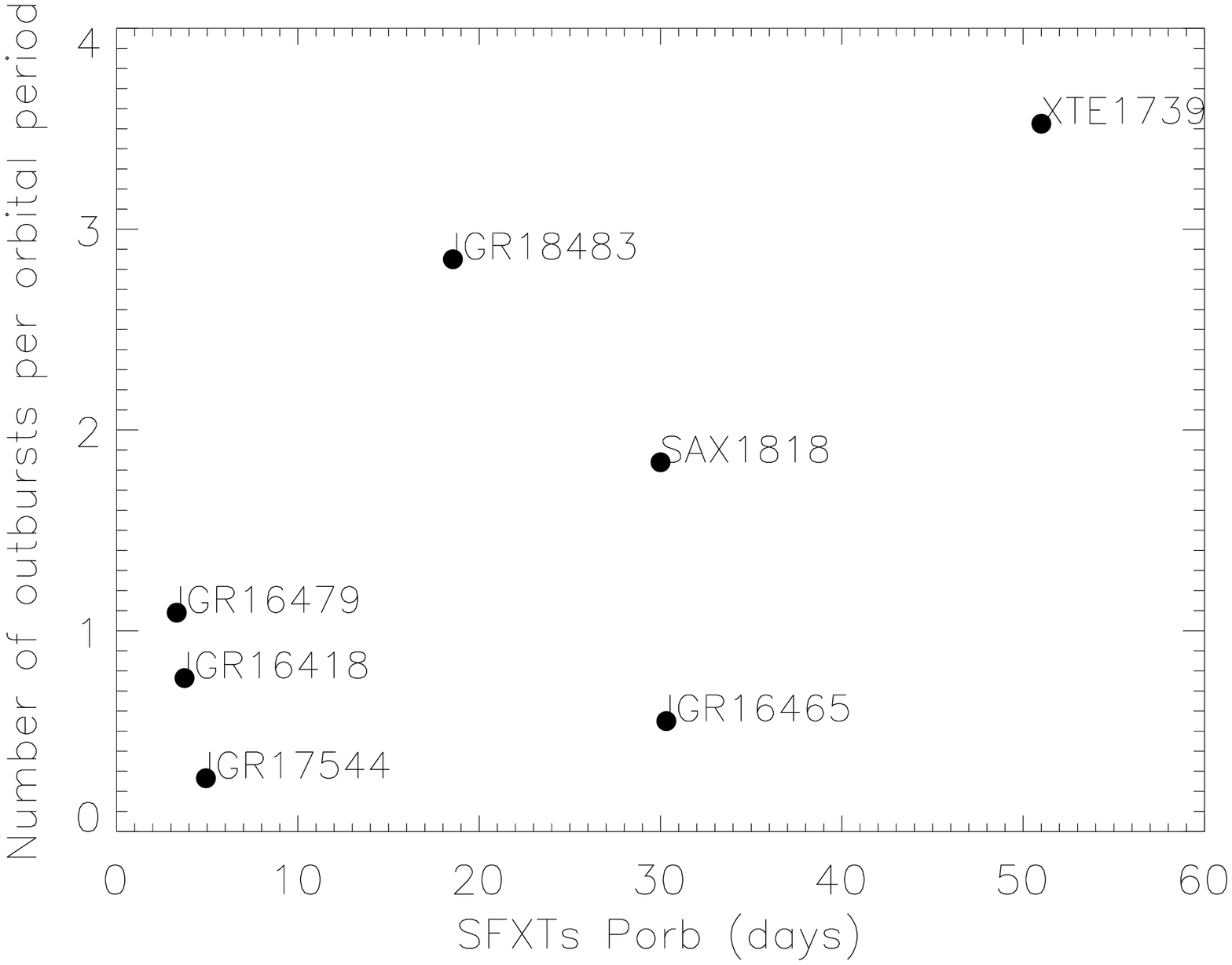} \\
\hspace{-1truecm}
\includegraphics[height=8.5cm,angle=-270]{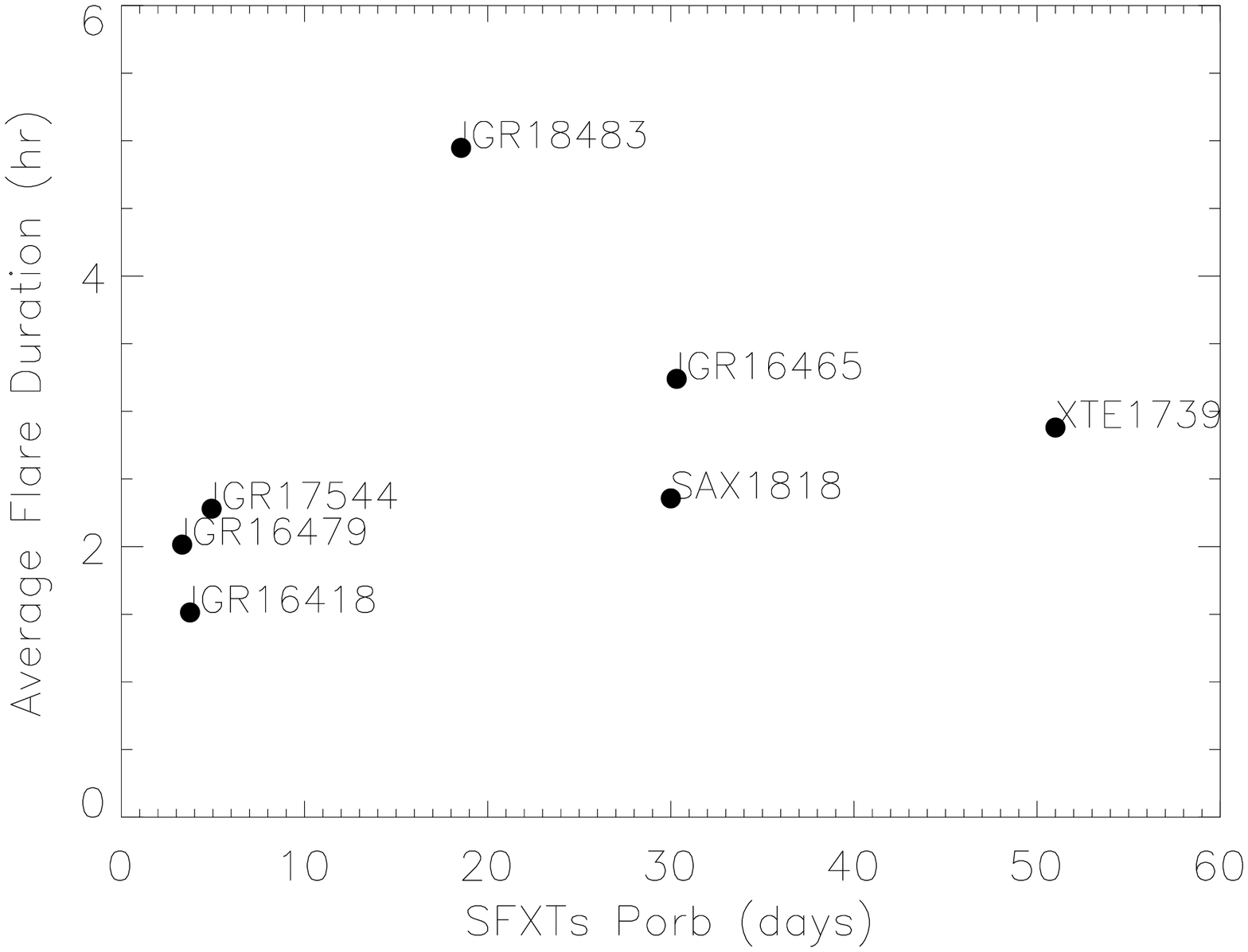} & 
\hspace{-1truecm}
\includegraphics[height=8.5cm,angle=-270]{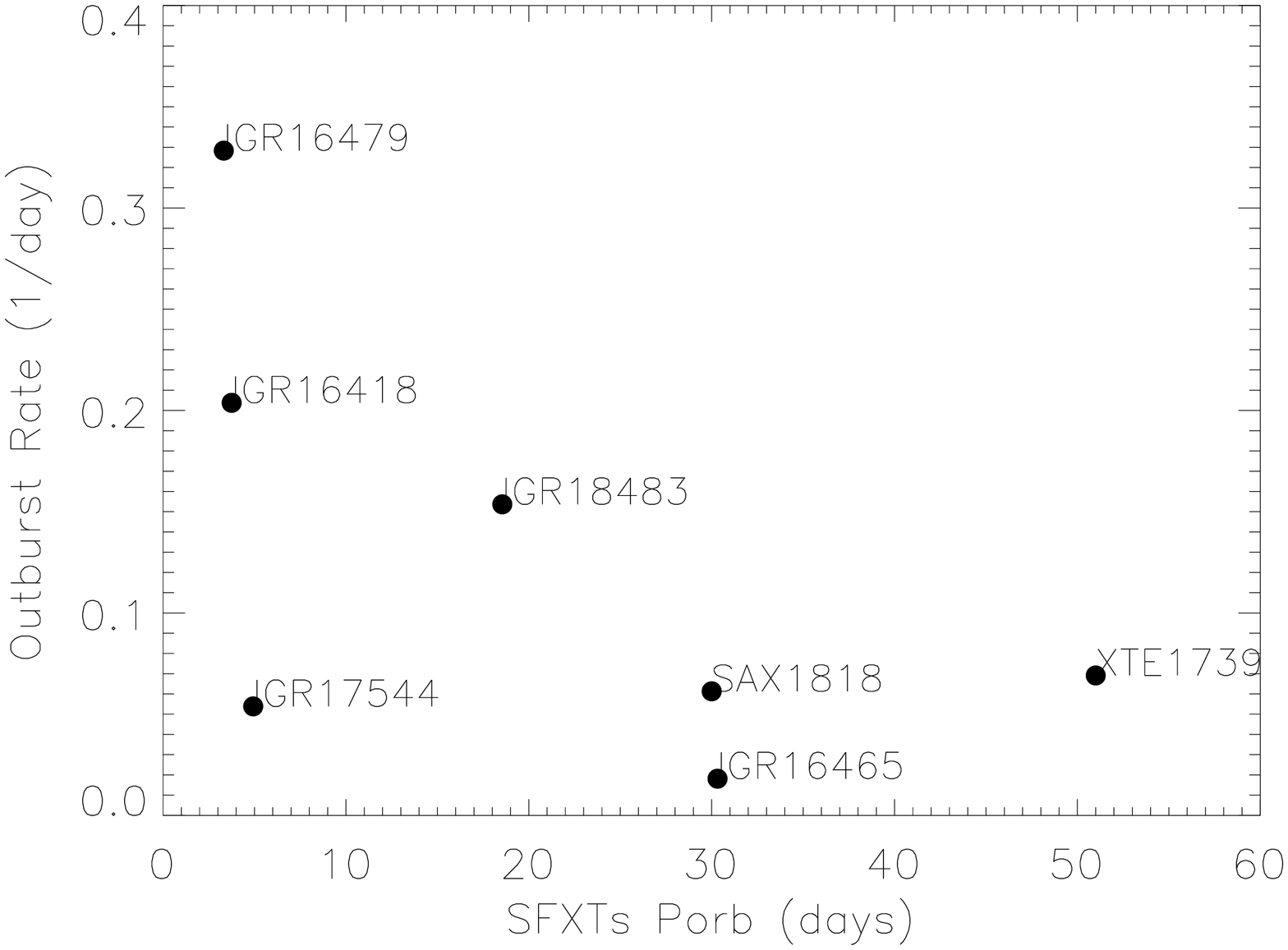} \\
\end{tabular}
\caption{\scriptsize $INTEGRAL$ archival data results for SFXTs with known orbital periods (IBIS data from Ducci, Sidoli \& Paizis, 2010).
Abbreviated source names are the following: 
IGR16479=IGRJ16479-4514, IGR16418=IGRJ16418-4532, IGR16465=IGRJ16465-4507, 
XTE1739=XTEJ1739-302, IGR17544=IGRJ17544-2619, IGR18483=IGRJ18483-0311, SAX1818=SAXJ1818.6-1703.
}
\label{lsfig:orb}
\end{figure*}
%

\begin{figure*}
\centering
\begin{tabular}{cc}
\hspace{-1truecm}
\includegraphics[height=8.5cm, angle=-270]{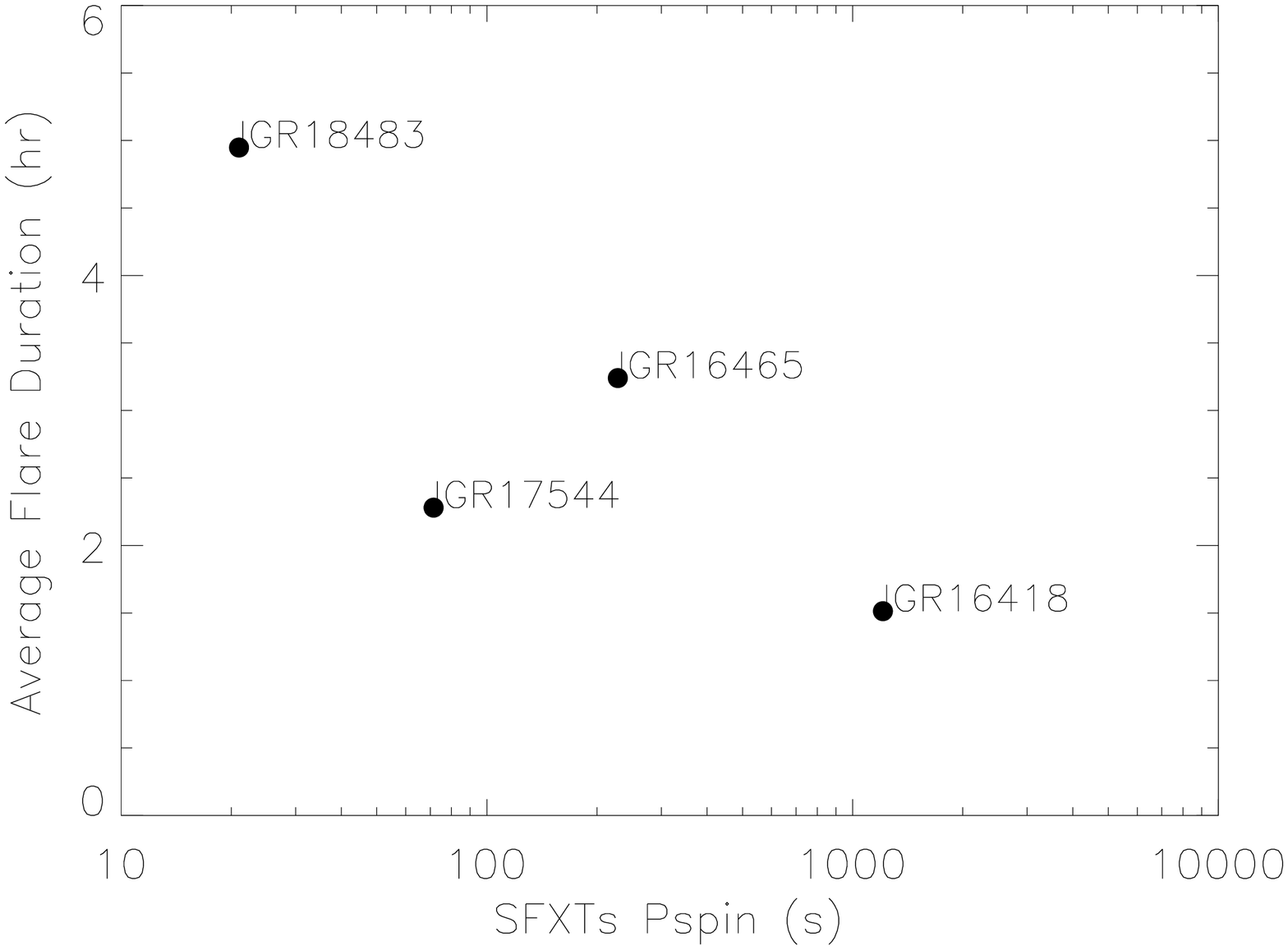} & 
\hspace{-1truecm}
\includegraphics[height=8.5cm, angle=-270]{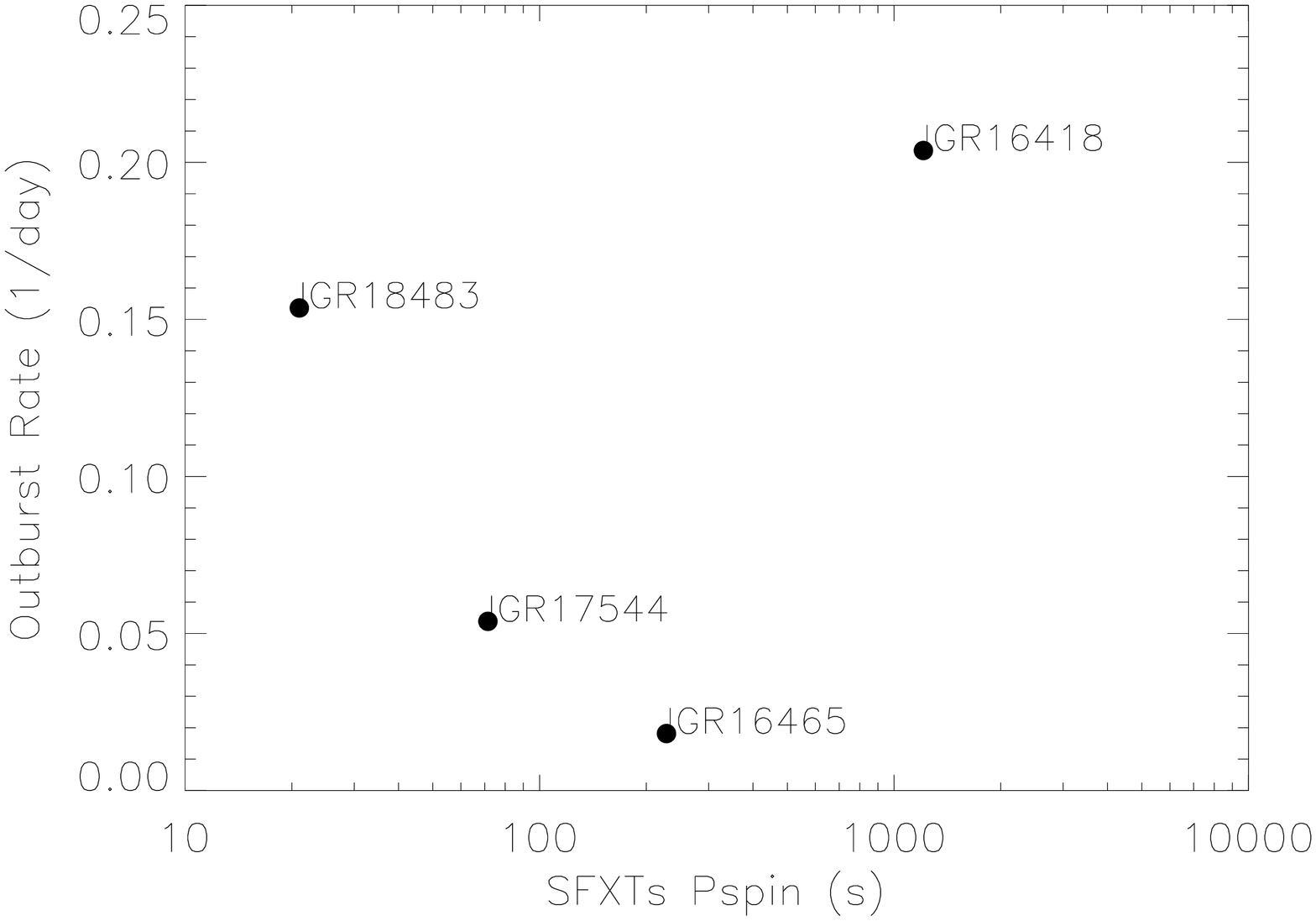} \\
\end{tabular}
\caption{\scriptsize $INTEGRAL$ archival data results for SFXTs with known spin periods (IBIS data are from Ducci, Sidoli \& Paizis, 2010). Abbreviated source names are the same as in Fig.~2.
}
\label{lsfig:spin}
\end{figure*}

Long-term properties of a sample of SFXTs, 
as observed with $INTEGRAL$/IBIS, are displayed 
in Fig.~\ref{lsfig:orb} versus the known orbital periods, and in Fig.~\ref{lsfig:spin} 
versus the spin periods (numbers have been calculated from 
data reported in Ducci et al. 2010): 
they are the percentage of time spent by SFXTs in bright flares,
the number of outbursts in a single orbital period, 
the average duration of the flaring activity 
and the outburst rate.
The trend for SFXTs with longer orbital periods is to show a lower outburst rate. 
This behavior agrees with the clumpy wind model (Negueruela et al. 2008).
On the other hand,
the SFXTs with the narrowest orbits can display very different outburst rates
(with IGRJ17544-2619 showing an outburst rate even lower than XTE~J1739--302, but with very different
orbital periods) demonstrating that 
the difference between  persistently accreting HMXB pulsars and SFXTs 
{\em cannot} be explained {\em only} by different orbital geometries.
Indeed, for example, the SFXTs IGR~J16479-4514 and IGR~J17544-2619 have orbital periods much shorter
than the bulk of persistently accreting, wind-fed pulsars (see also Fig.~1).

Besides the structure of the supergiant companion and the orbital configuration, some
authors considered a further possibility, that the compact object in SFXTs 
is slowly rotating ($\sim$1000 s) and 
that it undergoes a transition between the direct accretion regime and the onset of a centrifugal or a
magnetic barrier (Grebenev and Sunyaev 2007, Bozzo et al. 2008). 
In this latter case, the neutron star in SFXTs should be a magnetar (B$\sim$10$^{14}$--10$^{15}$~G).
These gated mechanisms, able to halt accretion rate most of the time, 
could explain the SFXTs with very short orbital periods where the compact object is always embedded 
within the dense and strong wind from the companion.

\section{\src: the SFXT with the shortest orbital period}

\begin{figure}[ht!]
\begin{center}
\vspace{0.truecm}
\includegraphics*[angle=-90,scale=0.45]{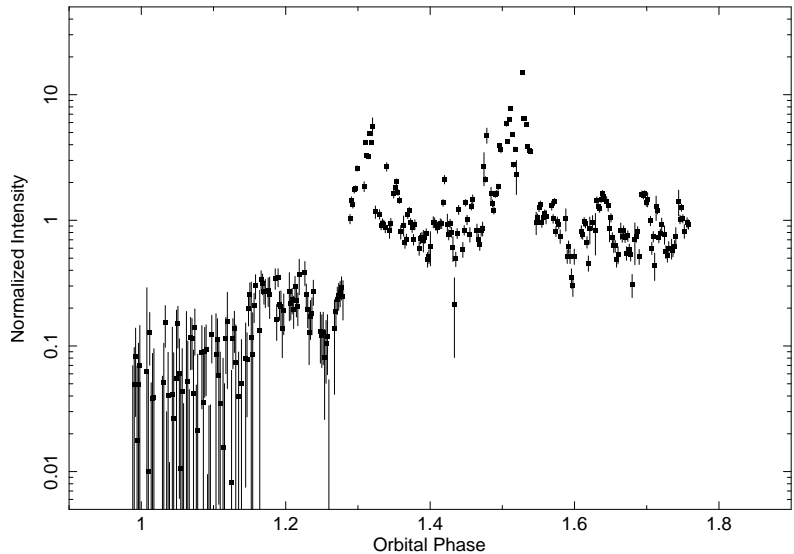}
\end{center}
\vspace{-0.85truecm}
\caption{\scriptsize IGR~J16479-4514 X--ray light curve  observed by $Suzaku$/XIS (1--10 keV) 
in February 2012 (Sidoli et al. 2013),
folded on the orbital period of 286792~s, assuming epoch 54547.05418 MJD.
}
\label{lsfig:suz_lc}
\end{figure}

In order to explore this possibility, to probe the X--ray and wind properties along
a single orbit in a SFXT, we obtained a 250~ks long $Suzaku$ observation of the
member with the shortest orbital period, IGR~J16479-4514 
(Sidoli et al. 2013, in press; Sidoli et al. these proceedings for details).
The XIS light curve (1--10 keV) observed in February 2012 is shown in Fig.~\ref{lsfig:suz_lc}, covering 
about 80\% of the orbit (P$_{orb}$=3.32~days).
%
%
The low X--ray intensity at the beginning of the observation 
is consistent with being due to the eclipse by the companion. 
Outside the eclipse, the source is highly variable, with two faint flares lasting 10--15~ks.  
The first flare is interestingly located at an orbital phase similar to other bright flares 
observed in the past (Bozzo et al. 2009), suggestive of the presence of a phase-locked gas stream
from the supergiant,
or wind component, which triggers an enhanced accretion of matter onto the compact object.
The average luminosity, assuming a distance of 2.8 kpc, is around 10$^{34}$~erg~s$^{-1}$.
The 1--10 keV time selected spectra can be well fit with an absorbed power law together with a 
narrow Fe~K${_\alpha}$ emission line at 6.4 keV. The intensity of the iron line is 
variable along the orbit and  correlates with the X--ray emission above 7 keV (outside the eclipse).
The resulting absorbing column density does not show evidence for variability,
(N$_{\rm H}$$\sim$10$^{23}$~cm$^{-2}$, in excess of the interstellar value), 
except from during the X--ray eclipse, where it is significantly lower, consistent with
the presence of Thomson scattering by electrons in the supergiant wind. 

The main result of the $Suzaku$ observation  
is that the scattered X--rays visible during the X--ray eclipse (compared with
the uneclipsed emission), 
allowed us to determine the  density of the donor wind at the orbital separation (a=2.2$\times$10$^{12}$~cm 
for a companion with M$_{opt}$=35~\msun)
resulting in 7$\times$10$^{-14}$~g~cm$^{-3}$. 
Assuming a spherical geometry for the outflowing supergiant wind, this density 
implies a ratio, $\dot{M}_{w}/ v_{\infty}$, between the wind mass loss rate
and the wind terminal velocity, of $7\times$10$^{-17}$~M$_{\odot}$/km.
This ratio,  
assuming  terminal velocities in the range from 500 to 3000~km~s$^{-1}$, translates into 
an accretion luminosity of  L$_{\rm X}$=3--15$\times$10$^{36}$~erg~s$^{-1}$, which is 
two orders of magnitude higher than that observed. 
In conclusion, 
a physical mechanism should  reduce the mass accretion rate.
A viable possibility is the mediating role of a neutron star magnetospheric surface.

\begin{figure*}
\centering
\begin{tabular}{cc}
\hspace{-1.0cm}
\includegraphics[height=8.0cm, angle=-90]{xmm_igr16479_timesel.ps} & 
\includegraphics[height=8.0cm, angle=-90]{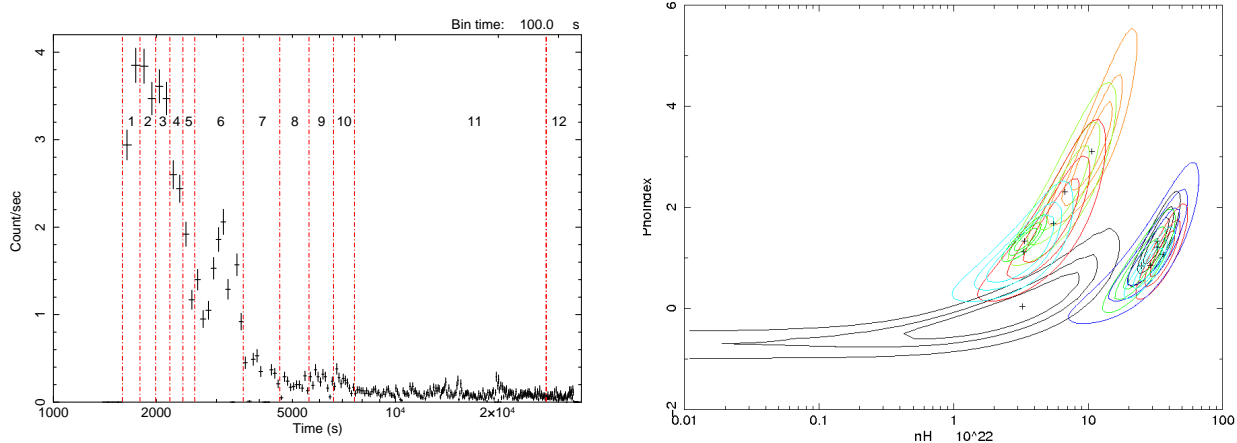} \\
\end{tabular}
\caption{\scriptsize  \emph{Left panel}: \xmm\ archival pn data of \src, observed on 2008, March 21 (I have
reanalysed obs. ID 0512180101, originally reported by Bozzo et al. 2008a). 
Numbers mark the time selected spectra during the X--ray eclipse ingress.
\emph{Right panel}: Confidence contour levels (68\%, 90\%, 99\%) 
for the two parameters (photon index and absorbing column density) 
of the absorbed power law fitting the \src\ EPIC pn time selected spectra 
(confidence contours of spectra 1--6 are concentrated around 3$\times$10$^{23}$~cm$^{-2}$, 
spectra 8--12 have contours around 6$\times$10$^{22}$~cm$^{-2}$, while spectrum numer ``7'' lies 
in between, with black contours).
}
\label{lsfig:xmm}
\end{figure*}

I have also reanalysed an archival $XMM-Newton$ observation performed during the 
eclipse ingress, originally reported by Bozzo et al. (2008a). 
I show the XMM light curve in Fig.~\ref{lsfig:xmm} (left panel) together
with numbers and vertical lines displaying the time intervals 
for the temporal selected spectra.
The results fitting the XMM spectra with an absorbed power law are displayed
on the right panel, where again the large variability in the absorption is only due to the
X--ray eclipse. Even with EPIC it is not possible to find any clear variability of the 
absorption (which would be expected in case of a clumpy wind), 
indicating a quite smooth nature for the \src\ supergiant wind, possibly because
we are probing here the inner regions of the companion wind.

In conclusion, in the SFXT \src\ the supergiant wind at the orbital separation is
too dense to explain the low X--ray luminosity and quite smooth to explain the source flares.

\acknowledgments
I would like to thank the organizers for their kind invitation 
at the  9th INTEGRAL Workshop ``An INTEGRAL view of the high-energy sky (the first 10 years)''
held in Paris, on 15-19 October 2012, to celebrate the 10th anniversary of the launch.
This work was supported in Italy by ASI-INAF contracts I/033/10/0 and I/009/10/0, and by 
the grant from PRIN-INAF 2009, ``The transient X--ray sky: 
new classes of X--ray binaries containing neutron stars''
(PI: L. Sidoli).


\begin{scriptsize}

\end{scriptsize}

\end{document}